\documentclass[aps,preprint,a4paper]{revtex4-2}
\usepackage{graphicx}
\usepackage{amsmath}
\usepackage{amssymb}
\usepackage{times}
\usepackage{ragged2e}
\usepackage[normalem]{ulem}
\usepackage[utf8]{inputenc}
\usepackage{placeins}
\usepackage[pdfborder={0 0 0}]{hyperref}
\hypersetup{
    colorlinks=true,
    linkcolor=blue,          
    citecolor=blue           
}

\def\beq{\begin{equation}}
\def\eeq{\end{equation}}
\def\beqna{\begin{eqnarray}}
\def\eeqna{\end{eqnarray}}
\def\bea{\begin{array}}
\def\ea{\end{array}}

\def\Re{\operatorname{Re}}
\def\Im{\operatorname{Im}}
\begin{document}
\title{Bifurcation-based phononic frequency combs in a forced van der Pol-Duffing oscillator}
\author{Adriano A. Batista$^1$, Madhurendra Mishra$^2$ and Adarsh Ganesan $^3$}
\affiliation{
$^1$ Departamento de Física, Universidade Federal de Campina Grande\\
Campina Grande-PB, CEP: 58109-970, Brazil\\
$^2$ Department of Physics, Sri Guru Tegh Bahadur Khalsa College, University of Delhi,\\
Arts Faculty Rd, University Enclave, New Delhi, Delhi, 110007, India\\
$^3$ Department of Electrical and Electronics Engineering, Birla Institute of Technology and Science (BITS) Pilani, Dubai Campus, \\Dubai International Academic City, Dubai, 345055, United Arab Emirates}
\date{\today}
\begin{abstract}
Here, we analyze the dynamics of a forced Van der Pol-Duffing oscillator in search of phononic frequency comb (PFC) generation.
When the drive amplitude is small, quasi-periodic self-oscillations are dominant, and we obtain a PFC spectrum response of the oscillator.
When the drive amplitude is sufficiently large, we obtain frequency locking.
We observe that near the transition to frequency locking the
PFC spectrum is strongest and has more peaks.
With increasing drive amplitude, the peaks coalesce, and the PFC spectrum disappears at an infinite-period saddle-node or homoclinic bifurcation.
We also study the stability of the frequency-locked responses and find
that a PFC spectrum can also occur near a subcritical Hopf bifurcation.
We find that the first-order averaged equations of our nonlinear dynamical
system are identical to the equations of motion of the Stuart-Landau oscillator.
We use averaging techniques up to second order to characterize the
quasi-periodic solutions.
Very good agreement with numerical integration results are obtained.
We complement our analysis by proposing a phenomenological model of nonlinear
dissipative coupling between two modes that accounts for the negative
dissipation term in the van der Pol-Duffing equation of motion.
\end{abstract}

\maketitle
\newpage
\section{Introduction}
\justify
Optical frequency combs have transformed precision measurement and spectroscopy \cite{Cundiff2003,Kippenberg2011}. In the frequency domain, they appear as a set of equally spaced, phase-coherent lines; in the time domain, the same superposition produces a periodic train of ultrashort pulses. Combs with closely spaced teeth arise from high-repetition-rate pulse trains, while those with coarser teeth come from longer-period oscillations. The metrological significance of this structure earned the 2005 Nobel Prize in Physics and prompted a sustained effort to carry the concept into wave domains beyond optics.
\justify
Phononic frequency comb (PFC) is the mechanical counterpart of optical frequency combs. A PFC is a spectrum of equidistant, phase-coherent vibrational lines, produced when phonon modes interact nonlinearly in a resonant structure. Its theoretical basis was established in 2014, when Cao et al. demonstrated through a Fermi-Pasta-Ulam framework that nonlinear resonances can self-organize vibrational energy into a comb-like distribution \cite{Cao2014}. The first experimental realization followed in 2017, when Ganesan, Do, and Seshia generated a PFC in a piezoelectric micromechanical resonator by exploiting three-wave mixing between a directly-driven mode and an autoparametrically excited sub-harmonic mode \cite{Ganesan2017}. Subsequent work from the same group showed that the comb structure persists across three-mode and multi-pathway coupling configurations, confirming that intermodal parametric coupling -- rather than any device-specific property -- is the generic mechanism \cite{Ganesan2018a,Ganesan2018b,Ganesan2018c}.

\justify
In the years that followed, PFC generation was reported
across a remarkably diverse set of platforms -- silicon
beam resonators \cite{czaplewski2018}, aluminium nitride
membranes \cite{Park2019}, graphene-silicon nitride
hybrids \cite{Singh2020}, bulk quartz crystals
\cite{Goryachev2020,Kubena2020}, molybdenum disulphide
membranes \cite{Chiout2021,Yousuf2023}, graphene drums
\cite{Keskekler2022}, silicon nitride strings and
membranes \cite{Ochs2022,deJong2023,Xiao2026},
rhombic \cite{Yang2021}, arch \cite{Gobat2023},
and ring \cite{Sun2023} micro-resonators, capacitive
CMOS-MEMS \cite{Bhosale2024}, carbon nanotube
\cite{Xu2025}, piezoelectric ultrasonic
\cite{kumar2025evolution}, nanocantilever NEMS
\cite{Mouharrar2024}, silicon carbide microdisk
\cite{Gou2025}, thin-film lithium niobate Lamb wave
\cite{Anderson2026}, and van der Waals ferromagnetic
\cite{Chen2025} resonators, and acoustically driven
bubble clusters \cite{Nguyen2021,Nguyen2023}.
The same physics has since surfaced in optomechanical
\cite{Zhang2021,Wang2024}, magnetomechanical
\cite{Vashahri2021,Xu2021}, and superconducting
electromechanical \cite{Han2022} systems, confirming
that the phenomenon is not tied to any particular
transduction modality. Theoretical work has extended
the framework to molecular rovibrational ladders
\cite{lei2025nonlinear}, twisted bilayer van der Waals
resonators \cite{liu2025phononic,arora2026twisted},
and bulk phonon transport
\cite{rangwala2026spontaneous}, while PFCs have
found use in underwater ranging \cite{Wu2019,Qian2022},
vibrational energy harvesting \cite{Bu2021,Zhang2023},
tunable sensing \cite{Ganesan2019sensing}, and
neuromorphic computing \cite{kumar2026neuromorphic},
accompanied by a growing portfolio of patents in
comb-enhanced oscillators, gradiometers, and signal
processors \cite{Kubena2021,Kubena2022a,Kubena2022b,
Kubena2023a,Kubena2023b,Song2023}.

\justify
PFC formation has long been associated with multi-mode interactions, yet a growing body of experimental work has shown that a single mechanical mode suffices. Ochs et al. \cite{Ochs2022} established this in a doubly clamped silicon nitride nanostring, where a monochromatic drive induced negative nonlinear friction that destabilized steady-state oscillation, producing regularly spaced sidebands without any inter-eigenmode frequency-matching condition. De Jong et al. \cite{deJong2023} demonstrated a complementary route by coupling a suspended dielectric membrane to a counter-propagating optical tweezer, where optothermal parametric pumping sustained self-oscillation while periodic optical gradient forces generated an overtone comb -- all with a milliwatt continuous-wave laser and no feedback or secondary mode. More recently, Xiao et al. \cite{Xiao2026} pushed single-mode generation into an ultrabroadband regime using an ultrathin silicon nitride membrane inside an optical cavity; above the phonon-lasing threshold, cascaded nonlinear mixing yielded a comb spanning up to 6000 teeth with spacing tunable across nearly four decades -- the broadest PFC reported to date.

\justify
Numerical studies of the van der Pol-Mathieu oscillator demonstrate that this minimal system supports tunable comb-like spectra. Ganesan et al. \cite{Ganesan2024a} showed that the comb spacing responds non-monotonically to a slowly varying perturbation of the resonance frequency -- decreasing initially, collapsing to a single spectral line across a finite dead-band, then re-emerging with reversed dependence -- with the dead-band width and branch sensitivities independently adjustable through drive amplitude and damping. A follow-up study demonstrated frequency synthesis: above the parametric resonance threshold the division ratio advances discretely through integer sub-multiples, while below it a fractional-ratio family reaching two orders of magnitude emerges, the two regimes separated by a discontinuous transition \cite{Ganesan2024b}. The drive amplitude thus serves as the primary handle in reconfiguring a single oscillator across negatively sensitive, positively sensitive, spectrally insensitive, and frequency-synthesizing states.

\justify
Complementary work has sought to ground frequency comb physics within the Duffing oscillator framework -- the standard model of MEMS/NEMS dynamics \cite{lifshitz2008,rhoads2010} -- encompassing bistability \cite{aldridge2005}, noise squeezing \cite{almog2006,almog2007noise}, coupled-resonator chaos \cite{karabalin2009}, nonlinear damping \cite{zaitsev2012}, and relaxation oscillations under two-tone excitation \cite{houri2019pwm,houri2020chaos}. On the comb generation front, Czaplewski et al. \cite{czaplewski2018} established that a saddle-node-on-invariant-circle bifurcation under single-tone drive is sufficient to seed a mechanical comb. Houri et al. \cite{houri2021librator} identified the ''librator'' regime, wherein Duffing nonlinearity sustains a multi-frequency limit cycle about rotating-frame fixed points; and Ochs et al. \cite{Ochs2022} showed that negative nonlinear friction alone suffices to generate a PFC. Batista and Lisboa de Souza \cite{batista2020} demonstrated, through an averaging analysis, that a pitchfork bifurcation delineates two comb regimes with distinct sideband spacings in a parametrically driven Duffing system. More recently, Mishra et al. \cite{mishra2025mems} showed that sufficiently strong nonlinearity drives comb dynamics into chaos, in another work, Mishra et al. \cite{mishra2026icnda} demonstrated through envelope analysis that cubic nonlinearity raises the Hopf threshold and progressively shrinks -- ultimately extinguishing -- the regime of stable PFC formation.

\justify
Shoshani and Shaw \cite{shoshani2023nonlinear} investigated frequency-comb generation through nonlinear interactions between modes with widely separated eigenfrequencies. Starting from a Hamiltonian model with quadratic intermodal coupling, they showed that a blue-detuned high-frequency mode can induce self-oscillations in a low-frequency mode. By eliminating the faster modal
dynamics, they obtained a forced Stuart--Landau equation
and subsequently used an Adler phase reduction to
describe injection locking and frequency pulling.
Their analysis established how these processes generate
tunable combs around both carrier frequencies, without
requiring conventional internal resonance.

\justify
Here, we take a complementary approach, starting from
an ac-forced van der Pol--Duffing oscillator whose
first-order averaged dynamics also reduce to the
Stuart--Landau equation. Rather than focusing on a
phase-only description of synchronization, we retain
the amplitude and phase dynamics to examine how
PFCs emerge and disappear through bifurcations. We derive analytical conditions for
saddle-node and Hopf bifurcations and identify three
routes to frequency locking: subcritical Hopf,
infinite-period saddle-node, and homoclinic
bifurcations. Second-order averaging improves the
prediction of the oscillation envelope and spectral
spacing, as confirmed by direct numerical integration.
We further propose a distinct physical realization
based on nonlinear dissipative coupling, in which
energy supplied to a lower-frequency mode is
transferred to a higher-frequency mode, producing
effective negative damping and self-oscillation. Together, these results connect the bifurcation structure of the Stuart--Landau oscillator to the onset, evolution, and eventual collapse of PFCs.

\justify
The remainder of this paper is organized as follows: in Sec.~\ref{sec:theory} we present our theoretical model and analyze it with the averaging method. In Sec.~\ref{sec:results} we present and discuss our analytical and numerical results. In Sec.~\ref{sec:conclusion} we draw our conclusions.

\section{Theory}
\label{sec:theory} 
The van der Pol-Duffing oscillator equation of motion in dimensionless format is
\beq
\ddot x= \gamma (1-x^2)\dot x-(1+\alpha x^2)x+F_\mathrm{d}\cos(\omega_\mathrm{d} t),
\label{eq:vdpDuffingOsc}
\eeq
where $\gamma$ is the dissipation coefficient, $\alpha$ is the Duffing
constant, $F_\mathrm{d}$ is the drive amplitude, 
$\omega_\mathrm{d}$ is the drive frequency.
In Appendix \ref{sec:modeCoupling}, we propose a phenomenological model of nonlinear dissipative coupling between two modes that accounts for the negative dissipation term in the above equation.

The bifurcations of the periodically forced van der Pol oscillator,
corresponding to the $\alpha=0$ limit of our model, were studied extensively by
Holmes and Rand using first-order averaging \cite{holmes1978bifurcations}. 
Here, we retain the Duffing nonlinearity and apply the averaging method
\cite{Guck83} to obtain the slow autonomous dynamics.
This is possible when $\gamma$, $\alpha$, and $F_\mathrm{d}$ are all $O(\epsilon)$,
where $0<\epsilon<<1$.
We transform to the slowly-varying variables $u(t)$ and $v(t)$ with the
transformation
\beq
\begin{aligned}
x(t) &=u(t)\cos(\omega_\mathrm{d} t)-v(t)\sin(\omega_\mathrm{d} t),\\
\dot x(t) &=-\omega_\mathrm{d}\left[u(t)\sin(\omega_\mathrm{d} t)+v(t)\cos(\omega_\mathrm{d} t)\right].
\end{aligned}
\label{eq:avgTrans}
\eeq
After applying this change of variables and neglecting fast oscillating terms, 
via Poincaré's weakly non-linear transformation, we obtain the averaged equations
\beq
\begin{aligned}
    \dot u &= \frac1{2
    \omega_\mathrm{d}}\left[ \gamma \omega_\mathrm{d} \left(1- \frac{r^2}4\right)u-\left(\Omega+\frac34
    \alpha r^{2}\right) v\right],\\
    \dot v &= \frac1{2 \omega_\mathrm{d}}\left[\gamma\omega_\mathrm{d}\left(1-\frac{r^2}4\right)v
    +\left(\Omega + \frac34\alpha r^2\right)u-F_\mathrm{d} \right]
    ,
\end{aligned}
\label{avgEqs}
\eeq
where 
$\Omega=1-\omega_\mathrm{d}^2$, and $r^2=u^2+v^2$.
In polar coordinates, $u=r\cos\theta$ and $v=r\sin\theta$, we obtain
\beq
\begin{aligned}
    \dot r &=
    \frac1{2\omega_\mathrm{d}}\left[\gamma\omega_\mathrm{d}\left(1-\frac{r^2}4\right)r-F_\mathrm{d}\sin\theta\right],\\
    \dot\theta &= \frac1{2\omega_\mathrm{d} r}\left[\left(\Omega+\frac{3\alpha}4
    r^2\right)r-F_\mathrm{d}\cos\theta\right].
\end{aligned}
\label{eq:polar}
\eeq
In complex coordinates, $z=u+iv=re^{\mathrm{i}\theta}$, we find
\beq
\dot z=\frac1{2\omega_\mathrm{d}}\left[\gamma\omega_\mathrm{d}\left(1-\frac{|z|^2}4\right)z
+i\left(\Omega+\frac{3\alpha}4|z|^2\right)z-iF_d\right].
\label{eq:stuartLandau}
\eeq
This is the equation of motion for the Stuart-Landau oscillator \cite{landau1944problem, Landau1965turbulence, Stuart1958}.
In the appendices {\bf B-D} we apply the averaging method to second-order approximation.
\subsection{Fixed-points of the Stuart-Landau equation}
When $F_\mathrm{d}=0$, the non-trivial fixed-point solutions of Eq.~\eqref{eq:stuartLandau} obey
\beq
\begin{aligned}
1-\frac{\bar r^2}4 &=0,\\
\Omega+\frac{3\alpha}4\bar r^2 &=0.
\end{aligned}
\label{charEq}
\eeq
In the first-order averaging approximation, the amplitude of these sustained
oscillations becomes $\bar r=2$ and the angular frequency is
$\omega_s=\sqrt{1+3\alpha}$.
When there is direct forcing, $F_\mathrm{d}\neq0$, we obtain
\beq
F_\mathrm{d}^2=\bar
r^2\left[\gamma^2\omega_\mathrm{d}^2\left(1-\frac{\bar
r^2}4\right)^2+\left(\Omega+\frac{3\alpha}4\bar r^2\right)^2\right].
\label{eq:roots}
\eeq
From Eq.~\eqref{eq:stuartLandau}, we find that the fixed-point solutions obey
\beq
\bar z =\bar u+i\bar v= 
\frac{iF_d}{\gamma\omega_\mathrm{d}\left(1-\frac{\bar r^2}4\right)
+i\left(\Omega+\frac{3\alpha}4\bar r^2\right)}.
\eeq
From this we find
\begin{equation}
\begin{aligned}
\bar u &= \frac{\left(\Omega+\frac{3\alpha}4\bar r^2\right)F_\mathrm{d}}
{\gamma^2\omega_\mathrm{d}^2\left(1-\frac{\bar r^2}4\right)^2
+\left(\Omega+\frac{3\alpha}4\bar r^2\right)^2},\\
\bar v &=
\frac{\gamma\omega_\mathrm{d}\left(1-\frac{\bar r^2}4\right)F_\mathrm{d}}
{\gamma^2\omega_\mathrm{d}^2\left(1-\frac{\bar r^2}4\right)^2
+\left(\Omega+\frac{3\alpha}4\bar r^2\right)^2}.
\end{aligned}
\end{equation}
From Eq.~\eqref{eq:roots}, we find that at a saddle-node bifurcation we have
\beq
\gamma^2\omega_\mathrm{d}^2\left(1-\frac{\bar r^2}4\right)^2+\left(\Omega+\frac{3\alpha}4\bar
r^2\right)^2+\bar r^2\left[
-\frac{\gamma^2\omega_\mathrm{d}^2}{2}\left(1-\frac{\bar r^2}4\right)+\frac{3\alpha}{2}\left(\Omega+\frac{3\alpha
\bar r^2}{4}\right)\right]=0.
\label{SNB}
\eeq
From this we find the following quadratic equation for $\bar r^2$ to be
\beq
\frac{3}{16}\left(\gamma^2\omega_\mathrm{d}^2+9\alpha^2\right)\bar r^4
+\left(3\alpha\Omega-\gamma^2\omega_\mathrm{d}^2\right)\bar r^2+\Omega^2
+\gamma^2\omega_\mathrm{d}^2=0,
\label{eq:detA=0}
\eeq
from where we find the solutions
\beq
\bar r^2_\pm=\frac{-\left(3\alpha\Omega-\gamma^2\omega_\mathrm{d}^2\right)\pm\sqrt{\left(3\alpha\Omega-\gamma^2\omega_\mathrm{d}^2\right)^2-
3/4\left(\gamma^2\omega_\mathrm{d}^2+9\alpha^2\right)\left(\Omega^2+\gamma^2\omega_\mathrm{d}^2\right)}}{3\left(\gamma^2\omega_\mathrm{d}^2+9\alpha^2\right)/8}.
\label{eq:r2_pm}
\eeq
\subsection{Limit-cycle solution}
We seek an approximate limit cycle solution of Eq.~\eqref{eq:stuartLandau} in
the form $z(t)=z_0+r_0e^{\mathrm{i}\delta_0 t}$, we find
\beq
\begin{aligned}
    0 &= (\gamma\omega_\mathrm{d}+\mathrm{i}\Omega)z_0+
    \frac{(-\gamma\omega_\mathrm{d}+3\mathrm{i}\alpha)}4\left(2r_0^2+|z_0|^2\right)z_0-iF_d,\\
    \mathrm{i}\delta_0 &=\frac1{2\omega_\mathrm{d}}\left[\gamma\omega_\mathrm{d}+\mathrm{i}\Omega+
    \frac{\left(-\gamma\omega_\mathrm{d}+3\mathrm{i}\alpha\right)}{4}\left(r_0^2+2|z_0|^2\right)\right].
\end{aligned}
\eeq
Using the first equation above, we solve for $z_0$
\beq
\begin{aligned}
z_0=&\frac{iF_d}{\gamma\omega_\mathrm{d}+\mathrm{i}\Omega+
    (-\gamma\omega_\mathrm{d}+3\mathrm{i}\alpha)\left(2r_0^2+|z_0|^2\right)/4}\\
    \implies&|z_0|^2=\frac{F_\mathrm{d}^2}{\gamma^2\omega_\mathrm{d}^2\left[1-\left(2r_0^2+|z_0|^2\right)/4\right]^2+\left[\Omega+\frac{3\alpha}{4}\left(2r_0^2+|z_0|^2\right)\right]^2}.
\end{aligned}
\eeq    
Taking the real and imaginary parts of the second equation above, we find
\beq
\begin{aligned}
4=&r_0^2+2|z_0|^2,\\
\delta_0=&\frac1{2\omega_\mathrm{d}}\left[\Omega+\frac{3\alpha}{4}\left(r_0^2+2|z_0|^2\right)\right]=\frac{\Omega+3\alpha}{2\omega_\mathrm{d}}=\frac{\omega_\mathrm{s}^2-\omega_\mathrm{d}^2}{2\omega_\mathrm{d}}.
\end{aligned}
\label{eq:limitCycle}
\eeq
From these, we can write the following equation for $|z_0|^2$
\beq
|z_0|^2=\frac{F_\mathrm{d}^2}{\gamma^2\omega_\mathrm{d}^2\left(1-3|z_0|^2/4\right)^2+\left[\Omega+3\alpha\left(2-\frac{3}{4}|z_0|^2\right)\right]^2}.
\eeq
We point out that these results are accurate only for very small values of $F_\mathrm{d}$.
\subsection{Stability analysis}
The linear stability of a fixed-point solution $\bar z$ of Eq.~\eqref{eq:stuartLandau} can be determined by the growth or decay of a small deviation $\Delta z$ from $\bar z$. 
We find
\beq
\Delta\dot z=\frac1{2\omega_\mathrm{d}}\left\{\gamma\omega_\mathrm{d}\left[ \left( 1-\frac{|\bar z|^2}{2}\right)\Delta z-\frac14\bar z^2\Delta z^*\right]+i\left[\left(\Omega+\frac{3\alpha}{2}|\bar z|^2\right)\Delta z+\frac{3\alpha}{4}\bar z^2\Delta z^*\right]\right\}.
\eeq
If we take the real and imaginary parts of $\Delta z=\Delta u+i\Delta v$,
we obtain
\beq
\begin{aligned}
    \Delta\dot u &=\frac1{2\omega_\mathrm{d}}\left\{\gamma\omega_\mathrm{d}\left( 1-\frac{\bar r^2}{2}\right)\Delta u-\left(\Omega+\frac{3\alpha}{2}\bar r^2\right)\Delta v-\frac{\gamma\omega_\mathrm{d}}4\Re\left(\bar z^2\Delta z^*\right)-
    \frac{3\alpha}{4}\Im\left(\bar z^2\Delta z^*\right)\right\},\\
    \Delta\dot v &=\frac1{2\omega_\mathrm{d}}\left\{\gamma\omega_\mathrm{d}\left( 1-\frac{\bar r^2}{2}\right)\Delta v+\left(\Omega+\frac{3\alpha}{2}\bar r^2\right)\Delta u-\frac{\gamma\omega_\mathrm{d}}4\Im\left(\bar z^2\Delta z^*\right)
    +\frac{3\alpha}{4}\Re\left(\bar z^2\Delta z^*\right)\right\}.
\end{aligned}
\eeq
With the help of the relations
\beq
\begin{aligned}
\Re\left(\bar z^2\Delta z^*\right) &= \left(\bar u^2-\bar v^2\right)\Delta u+2\bar u\bar v\Delta v,\\
\Im\left(\bar z^2\Delta z^*\right) &= 2\bar u\bar v\Delta u-\left(\bar u^2-\bar v^2\right)\Delta v,
\end{aligned}
\eeq
we write the linear differential equations for $\Delta u$ and $\Delta v$
\beq
\begin{aligned}
    \Delta\dot u &= \frac1{2
    \omega_\mathrm{d}}\left\{ \left[\gamma\omega_\mathrm{d}\left(1-
    \frac{\bar r^2}4-\frac{\bar u^2}2\right)-\frac{3\alpha}2 \bar u\bar v\right]\Delta
    u-\left[\Omega+\frac{3\alpha}4(\bar r^{2}+2\bar v^2)+\frac{\gamma\omega_\mathrm{d}}2\bar u\bar v\right]
    \Delta v\right\},\\
    \Delta\dot v &= \frac1{2
    \omega_\mathrm{d}}\left\{\left[\Omega + \frac{3\alpha}4 (\bar r^2+2\bar
    u^2)-\frac{\gamma\omega_\mathrm{d}}2\bar u\bar v\right]\Delta u+\left[\gamma\omega_\mathrm{d}\left(1-\frac{\bar r^2}4-\frac{\bar
    v^2}2\right)+\frac{3\alpha}2 \bar u\bar v\right]\Delta v
    \right\}.
\end{aligned}
\label{eq:linStab}
\eeq
From the characteristic equation, we find that the zero eigenvalue condition
is the same as Eq.~\eqref{eq:detA=0}, which occurs at a saddle-node bifurcation.
We also find that a necessary condition for a Hopf bifurcation to occur
is $\mathrm{Tr} J=\frac\gamma2\left(2-\bar r^2\right)=0\implies \bar
r=\sqrt{2}$, where $J$ is the Jacobian matrix of the linear ODE given in
Eq.~\eqref{eq:linStab}.
Replacing this in Eq~\eqref{eq:roots}, we find the Hopf bifurcation line
\beq
F_\mathrm{d}^2=
2\left[\frac{\gamma^2\omega_\mathrm{d}^2}4
+\left(\Omega+\frac{3\alpha}2\right)^2\right].
\label{eq:Hopf}
\eeq

\section{Results and discussion}
\label{sec:results}
\FloatBarrier
In this section, we discuss the dynamical properties of the forced van der Pol-Duffing oscillator.
Our objective is to numerically verify the validity of the main theoretical predictions of our model. More specifically, we want to map out the parameter space in which PFC generation is most intense in the belief that this will be a helpful guide for the experimentalist in its search.
We start by investigating the stability of the frequency-locked response and the SNBs that occur at certain values of the drive amplitude and the drive frequency.
We show that the averaging method captures the dynamics of this system very well.
We also show that the strongest PFC occurs near bifurcations such as subcritical Hopf and infinite-period bifurcations (saddle-node and homoclinic).

\begin{figure}[!ht]
    \centering{\includegraphics[scale=0.187]{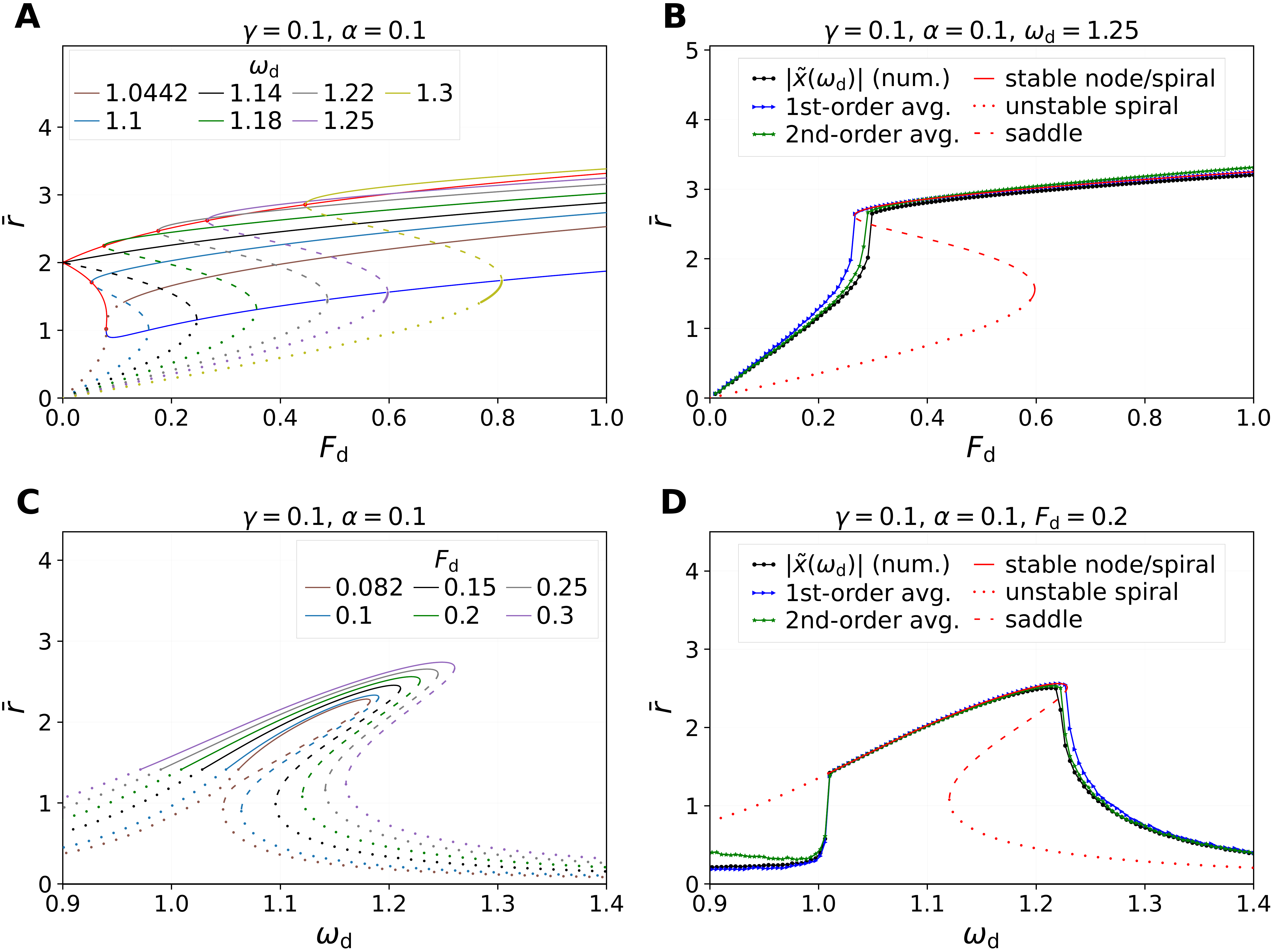}}
    \caption{{\bf A} Frequency-locked oscillation amplitudes as a function of drive
    amplitude $F_\mathrm{d}$ obtained from Eq.~\eqref{eq:roots} for various
    values of angular frequency $\omega_\mathrm{d}$. 
    The dotted line segments represent unstable spiral fixed points, the dashed
    line segments represent saddle fixed points, and the solid lines are stable
    spirals or nodes. The thin red and blue lines are the locus of saddle-node bifurcations given by Eq.~\eqref{eq:r2_pm}.
    {\bf B} 
The stationary amplitude of oscillations of the forced van der Pol-Duffing oscillator is given by the spectral component of the response of the oscillator
at $\omega_\mathrm{d}$.
The first-order averaging approximation is obtained from numerical integration of the Stuart-Landau equation given in Eq.~\eqref{eq:stuartLandau}.
The second-order averaging approximation is obtained from numerical integration
of Eq.~\eqref{eq:avg2forcedVDPolDuff2}.
In panel {\bf C} we plot the amplitude of the frequency-locked responses as function of the drive angular frequeny $\omega_\mathrm{d}$ of the
Stuart-Landau oscillator at various values of $F_\mathrm{d}$.
In panel {\bf D} we plot the frequency-locked response amplitude at $F_\mathrm{d}=0.2$ along with corresponding results obtained from the numerical integration of
the van der Pol-Duffing oscillator and from the averaging method.}  
    \label{fig:ampFreqLock}
\end{figure}

In Fig.~\ref{fig:ampFreqLock} {\bf A} we plot the frequency-locked oscillation
amplitude as a function of the drive amplitude that is implicitly given in Eq.~\eqref{eq:roots}.
We show  results for several drive frequencies at and around
$\omega_s=\sqrt{1+3\alpha}$, the angular frequency of the  self-oscillations of
the undriven van der Pol-Duffing oscillator.
For the parameter $\alpha$ shown on top of the panel, $\omega_\mathrm{s}\approx 1.14$. 
The dotted line branches are unstable spirals, i.e.~ the oscillator responses are not
frequency-locked, instead they become limit-cycle solutions.
For each value of $\omega_\mathrm{d}$, the locus of the two turning points of
Eq.~\eqref{eq:roots} is given in the analytical expression of $\bar r^2_\pm$ in Eq.~\eqref{eq:r2_pm}.
The thin red line represents $\bar r_+$ and the thin blue line represents 
$\bar r_-$.
This is an important result because it can guide the experimentalist in
finding the interesting parameter region where the PFC response is
strongest.
In panel {\bf B} we show the amplitude of the frequency-locked response of the
Stuart-Landau oscillator, Eq.~\eqref{eq:stuartLandau}, for $\omega_mathrm{d}$ as a function of $F_\mathrm{d}$ along with the numerical results obtained from the integration of the equations of motion given in Eq.~\eqref{eq:vdpDuffingOsc}.
A stationary long time series is Fourier transformed to obtain the amplitude of the spectral component at $\omega=\omega_\mathrm{d}$.
Similarly, we obtain this from the numerical integration of the Stuart-Landau
equation along with the inverse of the van der Pol transformation given in Eq.~\eqref{eq:avgTrans}.
This shows that for small values of $F_\mathrm{d}$, below the turning-point
value, the frequency locked solution is unstable and does not follow
Eq.~\eqref{eq:roots}.
In panel {\bf C} we show frequency-locked oscillation amplitudes as a function of
the angular drive frequency $\omega_d$ for various values of the drive amplitude.
There is a degenerate eigenvalue at the transition from the unstable spiral (dotted line) to a stable node fixed point (solid line), this occurs at $\bar r=\sqrt{2}$.
In panel {\bf D} we plot the frequency-locked solution along with the numerical
result for the Fourier transform amplitude at $\omega_d$.
On the left side there is a subcritical Hopf bifurcation, and on the right side there is homoclinic orbit bifurcation near the saddle-node bifurcation.

\begin{figure}[!ht]
    \centering{\includegraphics[scale=0.25]{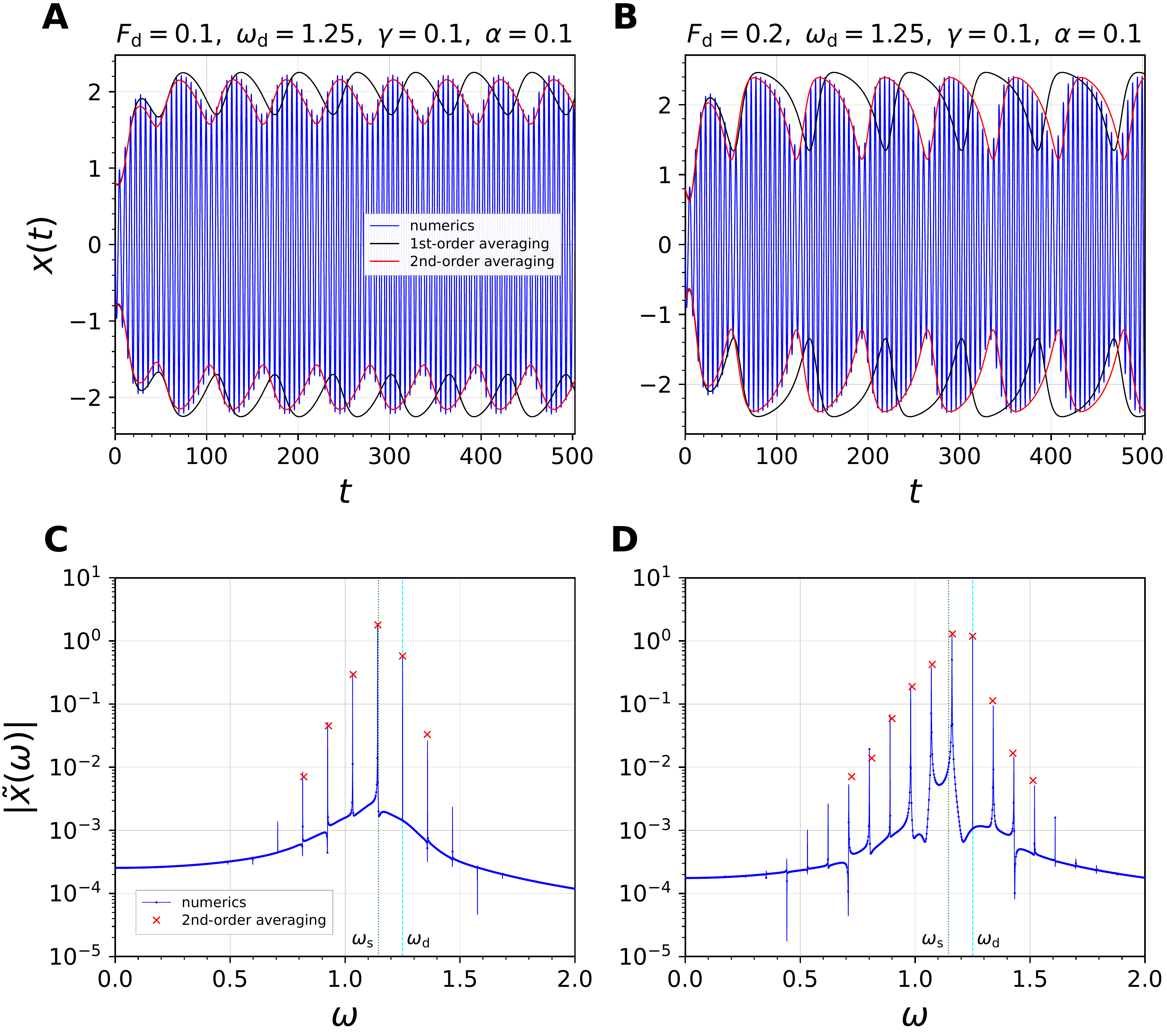}}
    \caption{PFC spectral response before the saddle-node bifurcation.
    Comparison between numerical results with the first and second-order averaging method results.
    {\bf A}. Numerical time series along the envelopes from the averaging method in first and second-order approximations. 
    {\bf C} Corresponding Fourier spectrum. The self oscillations angular frequency peak occurs at $\omega_\mathrm{s}\approx\omega_\mathrm{d}-\delta$ and the drive frequency at $\omega=\omega_\mathrm{d}$. The comb inter peak splitting is approximately $\delta$ given in Eq.~\eqref{eq:limitCycle}, which is an accurate result for small values of $F_\mathrm{d}$.
    Similar results are shown in panels {\bf C} and {\bf D}.
    Although the accuracy of the averaging method reduces with $F_\mathrm{d}=0.2$,
    the fitting is still very good.
    } 
    \label{fig:freqCombs} 
\end{figure}

In Fig.~\ref{fig:freqCombs} we show a cyclo-stationary solution
of the van der Pol-Duffing oscillator equation of motion \eqref{eq:vdpDuffingOsc}. 
In panel {\bf A}, one can see the numerical time-series curve (blue line) along with its envelope for $F_\mathrm{d}=0.1$.
The black line envelope is given by the first-order averaging approximation,
while the red line envelope is given by the second-order averaging approximation.
In panel {\bf C}, we show the corresponding spectral peaks of the time series in panel {\bf A}.
In panel {\bf B}, one can see the numerical time-series curve (blue line) along with its envelope for $F_\mathrm{d}=0.2$.
The humps get longer as the period of the limit-cycle becomes longer.
Consequently, more peaks appear in the Fourier transform spectrum characterizing
PFC generation as shown in panel {\bf D}.
Overall, very good agreement is obtained except for the large dephasing of the envelopes in the first-order averaging approximation.
This is due mainly to larger errors in estimating the period of the limit cycle
than in second-order averaging.

\begin{figure}[!ht]
    \centering{\includegraphics[scale=0.185]{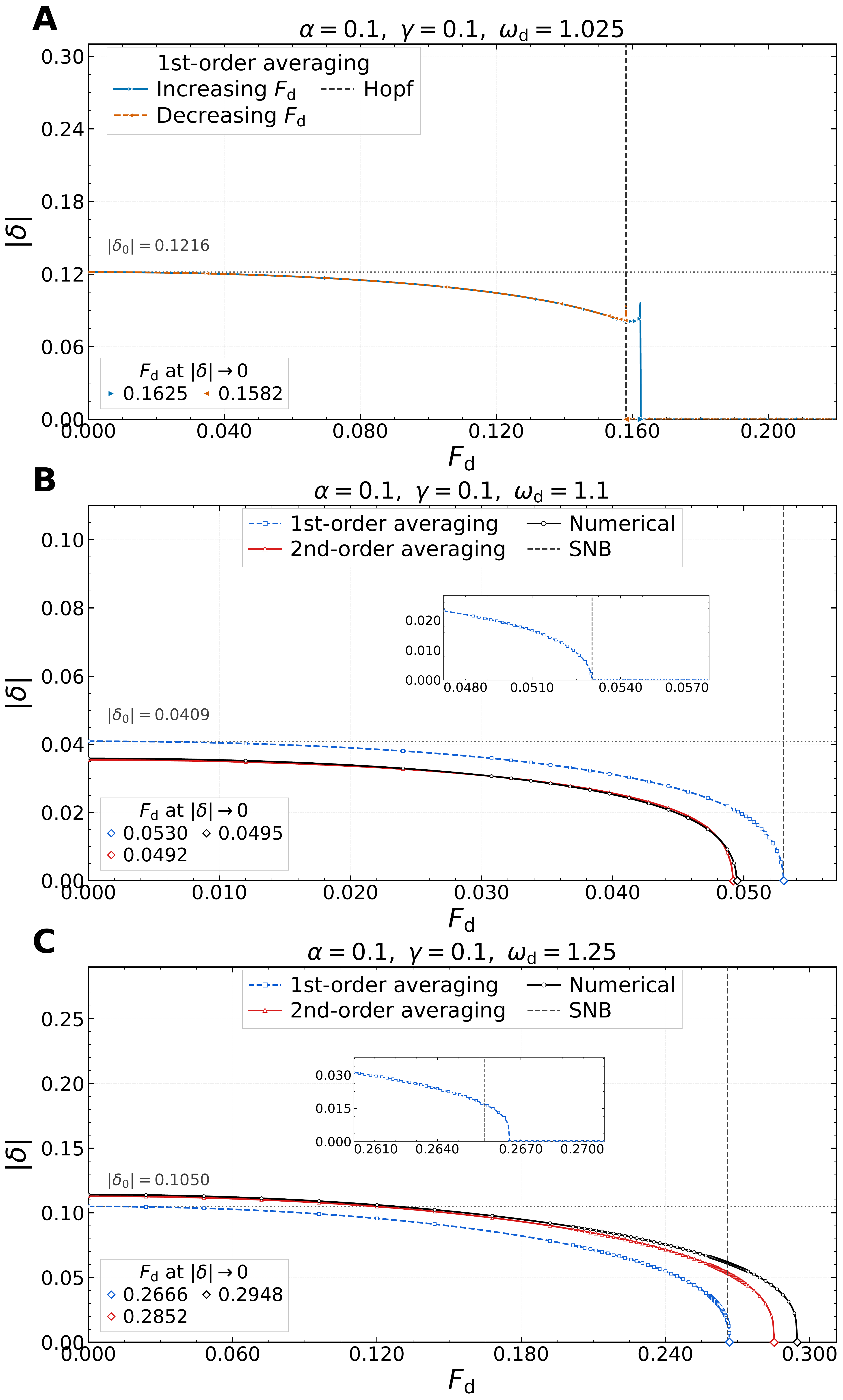}}
    \caption{PFC spectrum spacing ($\delta$) as a function of 
    drive amplitude $F_\mathrm{d}$. In panel {\bf A}, the comb disappears at
    a subcritical Hopf bifurcation. In panel {\bf B}, the comb disappears at an infinite-period saddle-node bifurcation. In panel {\bf C}, the comb disappears at an infinite-period homoclinic bifurcation. For the 1st-order averaging this occurs at
    $F_\mathrm{d}\approx 0.2666$, for the 2nd-order averaging approximation at $F_\mathrm{d}\approx0.2852$, and
    for the numerical integration this occurs at $F_\mathrm{d}\approx0.2948$.
    The insets zoom in at the infinite-period bifurcation points of the Stuart-Landau equation.} 
    \label{fig:combSpacing} 
\end{figure}

In Fig. \ref{fig:combSpacing}, we show the spectral spacing $\delta$, which is the spacing between consecutive spectral peaks, as a function of the drive amplitude $F_\mathrm{d}$. 
We show full numerical results of the integration alongside the
semi-analytical results from first- and second-order averaging.
In the first-order averaging approximation, we Fourier transform a stationary
long time series of $u(t)$ from the numerical integration of
Eq.~\eqref{avgEqs}.
Subsequently, we obtain $\delta$ from the peak spacing of the Fourier transform.
In the second-order approximation, we use Eqs.~\eqref{eq:avg2forcedVDPolDuff2} to obtain $u(t)$ and $v(t)$.
In the full numerical results, we obtain we Fourier transform a long stationary time-series of $x(t)$,
a solution of Eq. \eqref{eq:vdpDuffingOsc}, to obtain $\delta$.
In all panels, the horizontal dotted line indicates the analytical value of
$\delta$ for $F_d\rightarrow 0^+$ given in Eq.~\eqref{eq:limitCycle} and the
vertical dashed line indicates the saddle-node bifurcation given by replacing
$\bar r_+^2$ from Eq.~\eqref{eq:r2_pm} in Eq.~\eqref{eq:roots} to obtain
$F_\mathrm{d}$.
In panel {\bf A} we show the results for the peak splitting across a subcritical Hopf bifurcation. Due to this, we have a hysteresis cycle and overshoots are caused by the slow relaxation time at the bifurcation points.
In panel {\bf B} we show the peak splitting up to the infinite-period saddle-node bifurcation.
In panel {\bf C} we show the peak splitting up to the infinite-period homoclinic bifurcation, in which the limit-cycle becomes a homoclinic orbit (saddle connection) when $\delta$ goes to zero.
This global bifurcation does not coincide with the local SNB, as can be seen
in the inset figure, unlike the infinite-period SNB of panel {\bf B}.

\begin{figure}[!ht]
    \centering{\includegraphics[scale=0.20]{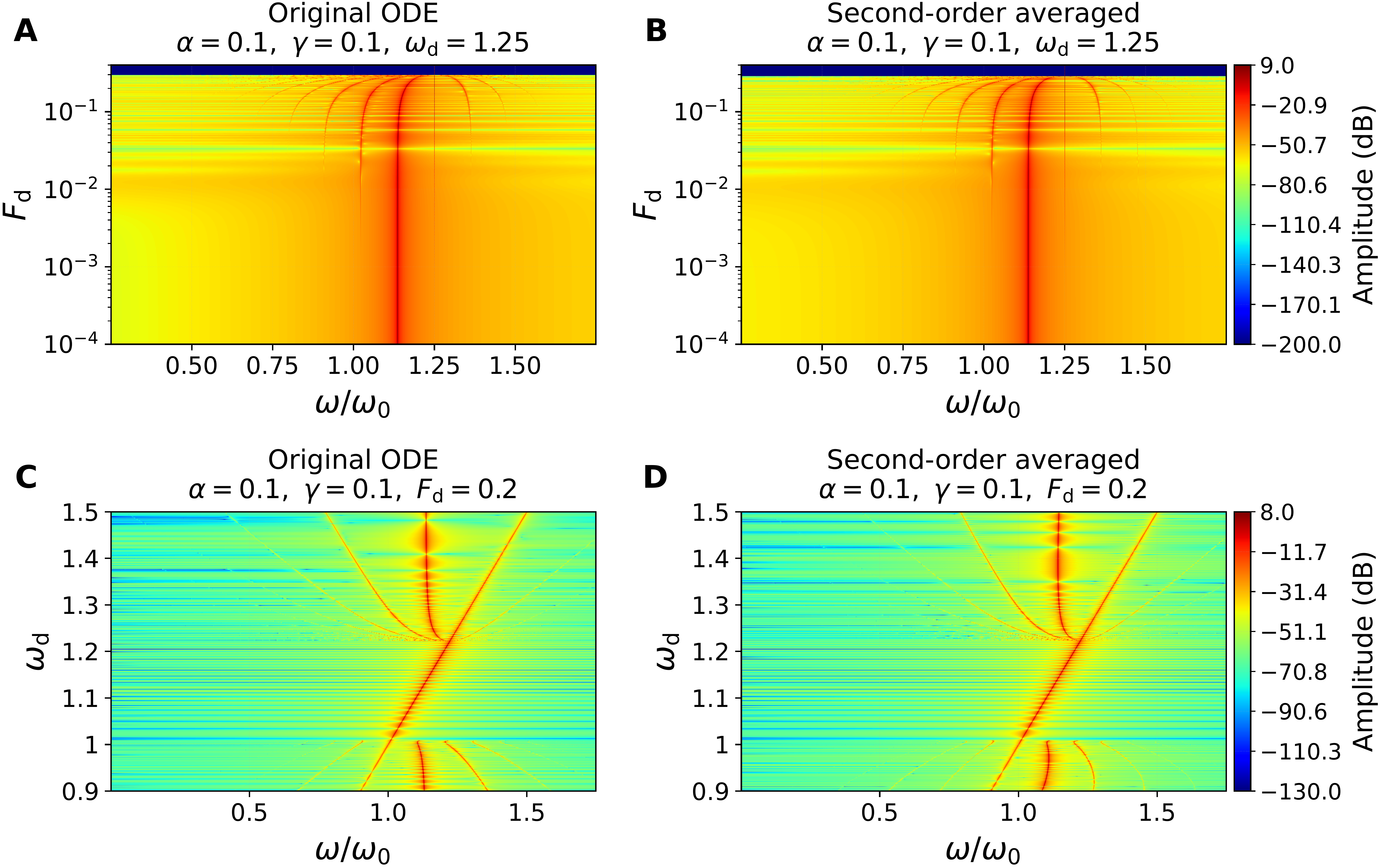}}
    \caption{
Spectral maps of the response of the van der Pol-Duffing oscillator to an the external
ac drive.
In panels {\bf A} and {\bf B}, we plot the frequency spectrum in terms of drive
amplitude $F_\mathrm{d}$ in log scale.
One sees the rise of the comb response as more and more peaks appear until the
collapse of the comb near $F_\mathrm{d}\approx 0.266$, where the period of
the limit cycle goes to infinity.
In panels {\bf C} and {\bf D}, we plot the frequency spectrum for each value of 
$\omega_\mathrm{d}$ along the vertical axis. 
The color code on the right side of panel {\bf D} indicates the amplitude of each peak in dB scale.
The abrupt collapse of the comb at $\omega_\mathrm{d}\approx1$ is due to
a subcritical Hopf bifurcation,
whereas the rise of the frequency comb at $\omega_\mathrm{d}\approx1.2$ is due to 
a bifurcation from a homoclinic orbit to a limit-cycle.
In this way, the PFC spacing rises continuously from zero
at the bifurcation point.
}
\label{fig:spectralMaps}
\end{figure}

In Fig.~\ref{fig:spectralMaps} we show spectral maps that help us visualize the onset and destruction of PFCs. We compare results from direct numerical integration of Eq.~\eqref{eq:vdpDuffingOsc}, depicted in the left column panels, along with results obtained from the integration of second-order averaged equations \eqref{eq:avg2forcedVDPolDuff2} shown in the right column panels.
The left and right panels present very similar results, attesting to the efficacy of the averaging method.
In panels {\bf A} and {\bf B}, we see that the comb collapses near $F_\mathrm{d}=0.266$ when the peak spacing decreases to zero. This collapse occurs at a homoclinic bifurcation. 
For larger values of $F_\mathrm{d}$, we obtain frequency locking.
In the lower panels {\bf C} and {\bf D}, we show how the spectrum of the oscillator response varies with the drive frequency $\omega_\mathrm{d}$. In these panels, we see two regions of comb generation (lower and higher values of $\omega_\mathrm{d}$)
intercalated with a region of frequency locking (in the middle values of 
$\omega_\mathrm{d}$).
This window of frequency locking is intuitively expected as 
$\omega_\mathrm{d}\approx\omega_\mathrm{s}$ in this middle region.
At $\omega_\mathrm{d}\approx 1$, the frequency comb ends at a subcritical Hopf bifurcation, while at $\omega_\mathrm{d}\approx1.23$, another frequency comb
arises in an infinite-period bifurcation as $\omega_\mathrm{d}$ increases.

\begin{figure}[!ht]
\centering{\includegraphics[scale=0.21]{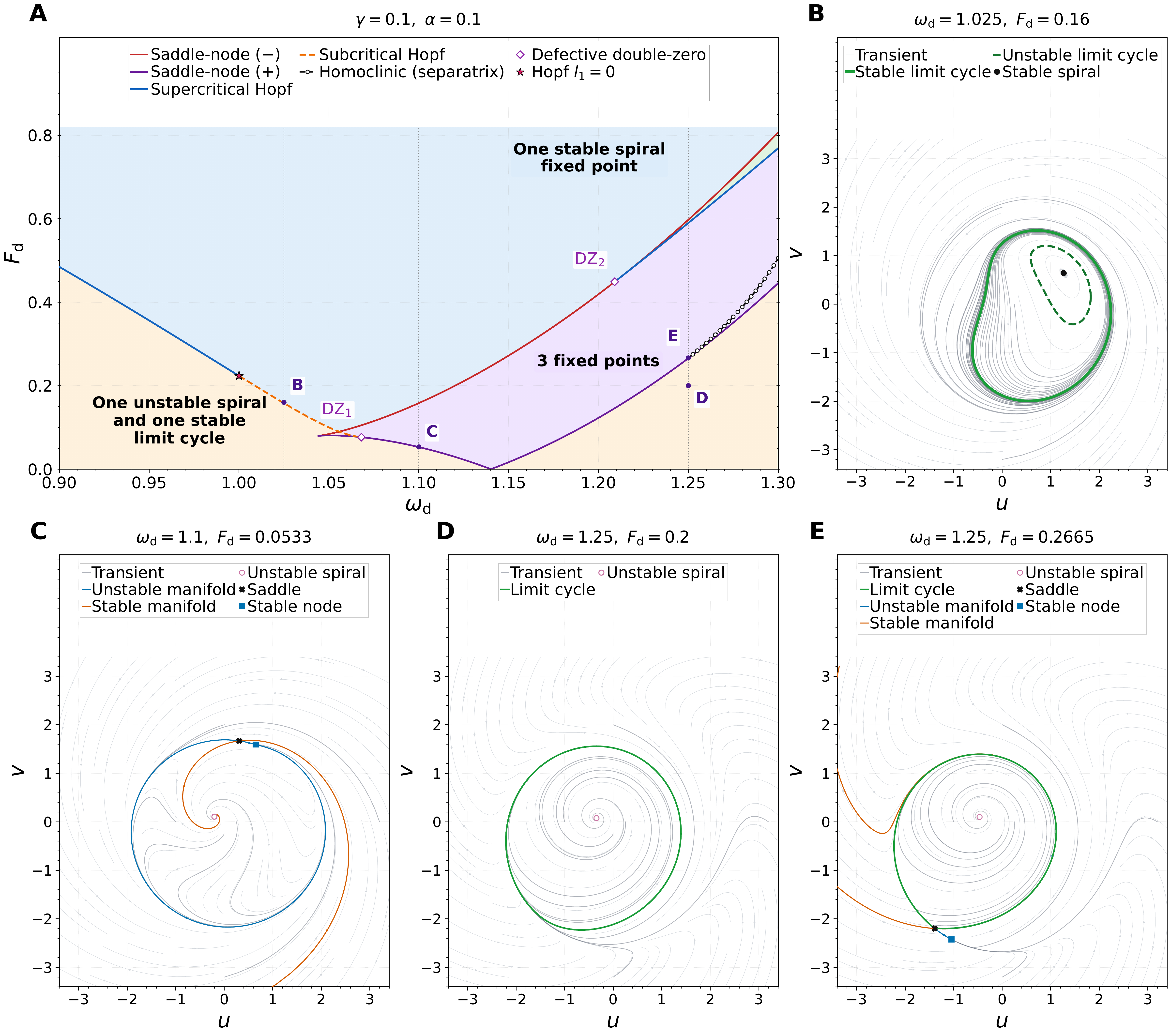}}
\caption{{\bf A} Bifurcation diagram of the Stuart-Landau oscillator in the $(\omega_\mathrm{d}, F_\mathrm{d})$ plane.
A supercritical Hopf bifurcation occurs on the solid blue line.
On the orange dashed-line a subcritical Hopf bifurcation occurs.
On the solid black and red lines a SNB occurs.
In the bluish region, we have a frequency-locked response, since there is only one stable spiral fixed point.
In the violet region, we have three fixed points: the unstable spiral, a saddle point, and a stable spiral or node.
In the beige region, we have an unstable spiral fixed point and a stable limit cycle.
In panel {\bf B}, one is slightly inside the bluish region near the subcritical Hopf bifurcation.
One has a stable spiral, an unstable limit cycle, and a stable limit cycle.
In panel {\bf C}, we represent the phase portrait with the saddle and stable
node fixed-point pair that appeared inside the limit cycle at an infinite-period
SNB.
{\bf D} A typical phase portrait of the Beige region with an unstable spiral and
a stable limit cycle..
{\bf E} Here, we depict the saddle connection in which the stable node is outside the homoclinic orbit.}
\label{fig:bif_diagram}
\end{figure}

In panel {\bf A} of Fig.~\ref{fig:bif_diagram} we show the bifurcation diagram
of the Stuart-Landau equation in the $(\omega_\mathrm{d}, F_\mathrm{d})$ space.
Local bifurcation lines are depicted as solid and dashed lines.
The dotted line is a global bifurcation.
We can distinguish three main domains of dynamical behavior.
In the beige region, we have one unstable spiral fixed-point and one stable limit cycle.
In this region, the dynamics is quasi-periodic.
This region is bounded by a supercritical Hopf bifurcation (solid blue line) on
the left side of the star point.
Along the dashed line, PFCs occur near a subcritical Hopf bifurcation.
To the right of the cusp point a SNB (solid violet line) occurs.
In the bluish region we have one stable spiral that corresponds to frequency locking.
The violet region is bounded by SNBs. Its lower boundary is near the region where
the strongest PFC spectrum occurs.
In one portion of it, approximately to the left of $\omega_\mathrm{d}=1.25$ down to the cusp point, an infinite-period SNB occurs. 
This global bifurcation appears simultaneously with the SNB.
The splitting between the infinite-period SNB and the homoclinic bifurcation (dotted line) becomes evident to the right of $\omega_\mathrm{d}=1.25$.
The strongest PFCs occur near these infinite-period bifurcations.
In the violet region, we could have up to three fixed-points:  two stable nodes
and one saddle or one unstable node.
We could also have a stable limit cycle whose amplitude is small, except in its lower region below the homoclinic orbit.
The upper boundary line of the violet region is given by another SNB (solid red line). 
In the phase portrait shown in panel {\bf B}, we have a stable limit cycle (in green), an unstable limit cycle (dashed line), and a stable spiral fixed point.
These occur near a subcritical Hopf bifurcation.
In panel {\bf C} we show two fixed points (a saddle and a stable node) that originate at an infinite-period SNB. Note that the unstable manifold of the saddle is the stable manifold of the node.
This occurred right after the limit cycle disappeared at this bifurcation.
There is also an unstable spiral inside these manifolds.
In panel {\bf D} we show an unstable spiral fixed point and a stable limit cycle
that is becoming oval in shape.
In panel {\bf E} we show a limit cycle that is very close to becoming a saddle connection at an infinite-period homoclinic bifurcation.
Naturally, there is a kink in the saddle connection, what does not appear in the
infinite period saddle-node bifurcation.
\FloatBarrier

\FloatBarrier
\section{Conclusion}
\label{sec:conclusion}
Here we investigated the dynamics of an ac forced van der Pol-Duffing
oscillator and found that it can be used as a new mechanism for PFC spectrum generation.
The interaction of the ac drive and the self-oscillations of this system can
lead to a response with a PFC spectrum.
We found that when the drive amplitude is small the self-oscillations dominate
and a quasi-periodic response happens, whereas when the ac drive is strong a frequency-locked response occurs. 
When both contributions are similar, PFC generation can occur.
We found that the averaged equations of motion of our model are given by
the dc forced Stuart-Landau equation, which has rich dynamics.
The main dynamical features of our oscillator are described by this equation.

For small drive amplitudes, the frequency-locked solution is unstable while
there is a stable limit-cycle solution.
As the drive amplitude is increased, this limit cycle ends at a Hopf bifurcation
or at an infinite-period bifurcation.
We observed three interesting bifurcations in which PFC generation becomes strong:
A subcritical Hopf bifurcation, an infinite-period SNB, and an infinite-period homoclinic bifurcation.
Near the subcritical Hopf bifurcation a frequency-comb-like response occurs, but
the spacing between the peaks is finite and the number of peaks is substantially
less than near one of the infinite-period bifurcations.
The period of the limit cycle increases until it diverges at an infinite-period bifurcation.
For lower values of drive frequency, this bifurcation is an infinite-period saddle-node bifurcation that occurs right on the local SNB.
For higher values of drive frequency, the limit cycle becomes a homoclinic orbit
at a homoclinic bifurcation.
This occurs when the unstable and stable manifolds of the saddle fixed point that arose at the SNB coincide. 
At this point, a kink appears in the previously smooth limit cycle.
After the limit cycle disappears, the frequency-locked solution becomes stable.
We found that the PFC generation is strongest near the onset of one of
these two global bifurcations.
We developed analytical expressions for the SNB and for a subcritical Hopf bifurcation.
We also point out that the homoclinic bifurcation does not coincide with the 
SNB, but occurs with higher values of drive amplitude.

The forced Van der Pol-Duffing oscillator could be implemented experimentally
in a dynamical system with nonlinear mode coupling in which most of the external energy is pumped in a high-frequency mode that is coupled to a low-frequency mode. 
One method of implementing this might be through the analytical model proposed in Ref.~\cite{shoshani2023nonlinear}.
Here, we developed a different nonlinear coupling between the modes that is
not based on the Hamiltonian of the system but rather on a dissipative coupling
between the modes that channels energy from the lower frequency mode into the
higher frequency mode (see Appendix \ref{sec:modeCoupling}).
Our method seems more intuitive and straightforward.
The energy flow could also be from the high frequency mode to the low frequency
mode, the conditions for this reverse flow are different from the low to high frequency flow direction.
We point out that in the case of weak turbulence in vibrating plates, the usual
flow direction is from low frequency to high frequency modes \cite{during2006weakTurbulence}.
This flow direction is very likely due to entropy production.
We mention that our mode coupling model predicts the onset of self-oscillations
only when the lower-frequency mode is pumped above a threshold.
This is qualitatively similar to the excitation of self oscillations in
the optomechanical instability observed in optical cavity mirrors due to radiation pressure backaction \cite{aspelmeyer2014cavity}.
A possible experimental implementation of this model could occur in a ring quantum cascade laser \cite{Piccardo2020} where a PFC generation was already observed. 
The underlying physics for a given cavity mode dynamics may be the Stuart-Landau
equation.
\appendix
\label{sec:appendix}
\section{Nonlinear mode coupling model}
\label{sec:modeCoupling}
We assume that two modes (or resonators) interact nonlinearly as follows
\beq
\begin{aligned}
m\frac{d^2 X}{ds^2}&= -\Gamma_0\frac{dX}{ds}+\beta\left(\frac{dY}{ds}\right)^2\frac{dX}{ds}-\eta X^2\frac{dX}{ds}-\left(k_0+\tilde\alpha X^2\right)X
+\tilde F_\mathrm{d}\cos\left(\tilde\omega_\mathrm{d} s\right),\\
M\frac{d^2 Y}{ds^2}&= -\Gamma_1\frac{dY}{ds}-\beta\left(\frac{dX}{ds}\right)^2\frac{dY}{ds}-k_1Y
+\tilde F_\mathrm{p}\cos\left(\tilde\omega_\mathrm{p} s\right), 
\label{eq:modeCoupling}
\end{aligned}
\eeq
where $m$ and $M$ are modal effective masses, 
$\Gamma_0$ and $\Gamma_1$ are modal dissipation coefficiants, $\beta$ is the nonlinear dissipation coupling constant, $k_0$ and $k_1$ are elastic constants, $\eta$ is the nonlinear dissipation constant, $\tilde\alpha$ is the Duffing constant, $\tilde F_\mathrm{p}$
is the pump amplitude,  $\tilde F_\mathrm{d}$ is the drive amplitude, 
$\tilde\omega_\mathrm{p}$ is the pump angular frequency, and 
$\tilde\omega_\mathrm{d}$ is the drive angular frequency.
The dissipative coupling between modes used above obeys energy conservation, i.e. the energy lost from the first mode due to this coupling is all transferred to the second mode.
If the mode with effective mass $M$ has substantially more dissipation than the second mode with effective mass $m$, we can neglect the effect of the second mode on the first mode.
In other words, we assume that $\Gamma_1>>\overline{\beta\left(\frac{dX}{ds}\right)^2}$.
In addition to that, when $M>>m$, the actuation of the smaller
resonator on the larger one is negligible.
Hence, we find the stationary response of the first mode to the pump, 
it is given approximately by
\beq
Y(s)=\frac{F_\mathrm{p}}2\left[\chi(\omega_\mathrm{p})e^{-i\omega_\mathrm{p} s}+\chi^*(\omega_\mathrm{p})e^{i\omega_\mathrm{p}s}\right],
\eeq
where $\chi(\omega)=\left(\omega_1^2-\omega^2+i\gamma\omega\right)^{-1}$
with $\omega_1^2=k_1/M$.
As we assume that $\omega_1<<\omega_0=\sqrt{k_0/m}$,
the contribution to the dynamics of the oscillating terms of $\left(\frac{dY}{ds}\right)^2$ can be neglected, as they are very off resonance. 
Consequently, we can approximate Eqs.~\eqref{eq:modeCoupling} by
\beq
m\frac{d^2 X}{ds^2}= -\Gamma_0\dot X+\beta\overline{\left(\frac{dY}{ds}\right)^2}\frac{dX}{ds}-\eta X^2\frac{dX}{ds}-\left(k_0+\tilde\alpha X^2\right)X
+\tilde F_\mathrm{d}\cos\left(\tilde\omega_\mathrm{d} s\right),
\label{eq:vdpDuffingOsc0}
\eeq
where the bar indicates a time average.
We find that
\[
\overline{\left(\frac{dY}{ds}\right)^2}=\frac{F_\mathrm{p}^2\omega_\mathrm{p}^2}2
|\chi(\omega_\mathrm{p})|^2.
\]
Replacing the above result in Eq.~\eqref{eq:vdpDuffingOsc0}, we obtain
\beq
m\frac{d^2 X}{ds^2}= \left(\Gamma-\eta X^2\right)\frac{dX}{ds}-\left(k_0+\tilde\alpha X^2\right)X
+\tilde F_\mathrm{d}\cos\left(\tilde\omega_\mathrm{d} s\right).
\label{eq:vdpDuffingOsc2}
\eeq
where $\Gamma=-\Gamma_0+\beta\frac{F_\mathrm{p}^2\omega_\mathrm{p}^2}2|\chi(\omega_\mathrm{p})|^2$ is the dissipation coefficient.
We notice that if the amplitude of oscillations of the first mode is small, then no self-oscillations occur since $\Gamma<0$.
On the other hand, if the amplitude of oscillations is above a threshold,
$\Gamma>0$, then
self-oscillations occur and the system dynamics can be approximated by a single
mode dynamics given by the ac driven van der Pol-Duffing oscillator.

We next convert Eq. \eqref{eq:vdpDuffingOsc2} to dimensionless format.
With this aim, we change the time variable to dimensionless time $t=\omega_0 s$ and divide the above equations by $m$. 
We find
\beq
\frac{d^2 X}{dt^2}= \left(\frac{\Gamma}{m\omega_0} -\frac{\eta}{m\omega_0} X^2\right)\frac{dX}{dt}-\left(1+\frac{\tilde\alpha}{m\omega_0^2} X^2\right)X
+\frac{\tilde F_\mathrm{d}}{m\omega_0^2}\cos\left(\frac{\tilde\omega_\mathrm{d}}{\omega_0} t\right),
\eeq
We then rewrite the above equation in dimensionless units with the
transformation $X=\kappa x$, where $\kappa=\sqrt{\frac{m\omega_0}\eta}$, 
$\gamma=\frac{\Gamma}{m\omega_0}$, $\alpha=\frac{\tilde\alpha\kappa^2}{m\omega_0^2}$,
$F_\mathrm{d}=\frac{\tilde F_\mathrm{d}}{m\kappa\omega_0^2}$, and 
$\omega_\mathrm{d}=\frac{\tilde\omega_\mathrm{d}}{\omega_0}$.
With all these changes, we find the equations of motion \eqref{eq:vdpDuffingOsc}
in dimensionless units.
This reduces the number of independent parameters and helps us connect the
experimental system to the theoretical model.
One could start the search for interesting nonlinear dynamics where PFCs occur with
parameters that could be experimentally realized.
This type of analysis could tell us which nonlinear phenomena are experimentally within reach.
\section{Averaging to second order: the van der Pol oscillator}
The averaged equations to second-order of the van der Pol oscillator (where $F_\mathrm{d}=0$ and $\alpha=0$) are given by
\beq
\begin{aligned}
    \dot u &= \frac1{2\omega_\mathrm{s}}
    \left[ \gamma \omega_\mathrm{s} \left(1- \frac{r^2}4\right)u-\Omega v\right]+\Delta u_2,\\
    \dot v &= \frac1{2 \omega_\mathrm{s}}
    \left[\gamma\omega_\mathrm{s}\left(1-\frac{r^2}4\right)v+\Omega u\right]+\Delta v_2
 ,
\end{aligned}
\eeq
where the second-order averaging corrections are given by
\beq
\begin{aligned}
    \Delta u_2&={\frac {11}{256}}\,{\frac {{\gamma}^{2}{u}^{4}v}{\omega_\mathrm{s}}}+\frac1{16}\,{
\frac {\gamma\,\Omega\,{u}^{3}}{{\omega_\mathrm{s}}^{2}}}+ \left( {\frac {11}{128
}}\,{\frac {{\gamma}^{2}{v}^{3}}{\omega_\mathrm{s}}}-\frac3{16}\,{\frac {{\gamma}^{2}v}
{\omega_\mathrm{s}}} \right) {u}^{2}+\frac1{16}\,{\frac {\gamma\,{v}^{2}u\Omega}{{
\omega_\mathrm{s}}^{2}}}+{\frac {11}{256}}\,{\frac {{\gamma}^{2}{v}^{5}}{\omega_\mathrm{s}}}\\
&-\frac3{16}\,{\frac {{\gamma}^{2}{v}^{3}}{\omega_\mathrm{s}}}+{\frac {1}{256}}\,{\frac 
{ \left( 32\,{\Omega}^{2}+32\,{\gamma}^{2}{\omega_\mathrm{s}}^{2} \right) v}{{
\omega_\mathrm{s}}^{3}}}\\
\Delta v_2&=-{\frac {11}{256}}\,{\frac {{\gamma}^{2}{u}^{5}}{\omega_\mathrm{s}}}+ \left( -{
\frac {11}{128}}\,{\frac {{\gamma}^{2}{v}^{2}}{\omega_\mathrm{s}}}+\frac3{16}\,{\frac {
{\gamma}^{2}}{\omega_\mathrm{s}}} \right) {u}^{3}+\frac1{16}\,{\frac {\gamma\,\Omega\,v
{u}^{2}}{{\omega_\mathrm{s}}^{2}}}\\
&+ \left( -{\frac {11}{256}}\,{\frac {{\gamma}^{
2}{v}^{4}}{\omega_\mathrm{s}}}+\frac{3}{16}\,{\frac {{\gamma}^{2}{v}^{2}}{\omega_\mathrm{s}}}-{
\frac {1}{256}}\,{\frac {32\,{\Omega}^{2}+32\,{\gamma}^{2}{\omega_\mathrm{s}}^{2}
}{{\omega_\mathrm{s}}^{3}}} \right) u+\frac1{16}\,{\frac {\gamma\,\Omega\,{v}^{3}}{{
\omega_\mathrm{s}}^{2}}}
\end{aligned}
\eeq
This can be simplified to
\beq
\begin{aligned}
    \Delta u_2&=\frac{\gamma\Omega r^2}{16\omega_\mathrm{s}^2}u
    +\frac1{8\omega_\mathrm{s}}\left[\frac{11\gamma^2r^4}{32}
            -\frac{3\gamma^2r^2}{2}
            +\frac{(\Omega^2+\gamma^2\omega_\mathrm{s}^2)}{\omega_\mathrm{s}^2}\right]v\\
\Delta v_2&=-\frac1{8\omega_\mathrm{s}}\left[\frac{11\gamma^2r^4}{32}
                    -\frac{3\gamma^2r^2}{2}
                    +\frac{(\Omega^2+\gamma^2\omega_\mathrm{s}^2)}{\omega_\mathrm{s}^2}\right]u
            +\frac{\gamma\Omega r^2}{16\omega_\mathrm{s}^2}v
\end{aligned}
\eeq
\beq
\begin{aligned}
    \dot u &= \frac1{2\omega_\mathrm{s}}
    \left\{ \gamma \omega_\mathrm{s} \left[1-
    \frac{r^2}4\left(1-\frac\Omega{2\omega_\mathrm{s}^2}\right)\right]u-\tilde\Omega
    v\right\},\\
    \dot v &= \frac1{2 \omega_\mathrm{s}}
    \left\{\tilde\Omega
    u+\gamma\omega_\mathrm{s}\left[1-\frac{r^2}4\left(1-\frac\Omega{2\omega_\mathrm{s}^2}\right)\right]v\right\}
 ,
\end{aligned}
\eeq
in which the renormalized detuning is given by
\[
\tilde\Omega =\Omega-\frac14\left[\frac{11\gamma^2r^4}{32}-\frac{3\gamma^2r^2}2
+\frac{\Omega^2+\gamma^2\omega_\mathrm{s}^2}{\omega_\mathrm{s}^2}\right].
\]
This leads to self-sustained oscillations with amplitude and frequency determined by the following equations
\beq
\begin{aligned}
    \bar r^2 &= \frac4{1-\frac\Omega{2\omega_\mathrm{s}^2}},\\
    \Omega &=
    \frac14\left[\frac{11\gamma^2}{2(1-\Omega/2\omega_\mathrm{s}^2)^2}
    -\frac{6\gamma^2}{1-\Omega/2\omega_\mathrm{s}^2}+\frac{\Omega^2+\gamma^2\omega_\mathrm{s}^2}{\omega_\mathrm{s}^2}\right],
\end{aligned}
\eeq
where here $\Omega=1-\omega_\mathrm{s}^2$.
One can then solve for $\omega_\mathrm{s}$ and then find $\bar r$.
\section{Averaging to second order: the van der Pol-Duffing oscillator}
\beq
\begin{aligned}
    \dot u &= \frac1{2
    \omega_\mathrm{s}}\left[ \gamma \omega_\mathrm{s} \left(1- \frac{r^2}4\right)u-\left(\Omega+\frac34
    \alpha r^{2}\right) v\right]+\Delta u_2,\\
    \dot v &= \frac1{2 \omega_\mathrm{s}}\left[\gamma\omega_\mathrm{s}\left(1-\frac{r^2}4\right)v
    +\left(\Omega + \frac34\alpha r^2\right)u\right]+\Delta
    v_2
    ,
\end{aligned}
\label{avg2Eqs}
\eeq
\beq
\begin{aligned}
    \Delta u_2&=\frac1{32}{\frac {\alpha\gamma\,{u}^{5}}{{\omega_\mathrm{s}}^{2}}}-{\frac {1}{256}}\,
{\frac { \left( -11\,{\gamma}^{2}{\omega_\mathrm{s}}^{2}-51\,{\alpha}^{2}
\right) v{u}^{4}}{{\omega_\mathrm{s}}^{3}}}+ \frac1{16}\left( {\frac {\alpha\,\gamma
\,{v}^{2}}{{\omega_\mathrm{s}}^{2}}}+{\frac {\gamma\,\Omega}{{\omega_\mathrm{s}}^{2}}}
 \right) {u}^{3}\\
 &-\frac {1}{256\omega_\mathrm{s}^3}\left[\left( -22\,{
\gamma}^{2}{\omega_\mathrm{s}}^{2}-102\,{\alpha}^{2} \right) {v}^{3}
+\left( -96\,\alpha\,\Omega+48\,{\gamma}^{2}{\omega_\mathrm{s}}^{2} \right) v
\right] {u}^{2}\\
&+
 \left[ 1/32{\frac {\alpha\,\gamma\,{v}^{4}}{{\omega_\mathrm{s}}^{2}}}+1/16\,{
\frac {\gamma\,\Omega\,{v}^{2}}{{\omega_\mathrm{s}}^{2}}}
\right] u\\
&-{\frac {1}{256}}\,{\frac { \left( -11\,{\gamma}^{2}{
\omega_\mathrm{s}}^{2}-51\,{\alpha}^{2} \right) {v}^{5}}{{\omega_\mathrm{s}}^{3}}}
-{\frac {1
}{256}}\,{\frac { \left( -96\,\alpha\,\Omega+48\,{\gamma}^{2}{\omega_\mathrm{s}}^
{2} \right) {v}^{3}}{{\omega_\mathrm{s}}^{3}}}\\
&+{
\frac {1}{8}}\,{\frac { \left( {\Omega}^{2}+{\gamma}^{2}{
\omega_\mathrm{s}}^{2} \right) v}{{\omega_\mathrm{s}}^{3}}}\\
\Delta v_2 &=-{\frac {1}{256}}\,{\frac { \left( 51\,{\alpha}^{2}+11\,{\gamma}^{2}{
\omega_\mathrm{s}}^{2} \right) {u}^{5}}{{\omega_\mathrm{s}}^{3}}}+1/32\,{\frac {\gamma\,v
\alpha\,{u}^{4}}{{\omega_\mathrm{s}}^{2}}}\\
&-\frac {1}{256\omega_\mathrm{s}^3}\,\left( 
 \left( 102\,{\alpha}^{2}+22\,{\gamma}^{2}\omega_\mathrm{s}^2 \right) {v}^{2}
+96\,\alpha\,\Omega-48{\gamma}^{2}{\omega_\mathrm{s}}^{2} \right) {u}^{3}\\
& + \frac1{16}\,\left( {
\frac {\alpha\,\gamma\,{v}^{3}}{{\omega_\mathrm{s}}^{2}}}+{\frac {\Omega\,
\gamma\,v}{{\omega_\mathrm{s}}^{2}}}\right) {u}^{2}\\
 -&{\frac {1}{256}}\left[{\frac { \left( 51\,{\alpha}^{2}+11\,{\gamma}^{2}{
\omega_\mathrm{s}}^{2} \right) {v}^{4}}{{\omega_\mathrm{s}}^{3}}}+{\frac {
 48\left(2\alpha\,\Omega-{\gamma}^{2}{\omega_\mathrm{s}}^{2} \right) {v}^{2}
}{{\omega_\mathrm{s}}^{3}}}
+{\frac 
{32({\gamma}^{2}{\omega_\mathrm{s}}^{2}+{\Omega}^{2})}{{\omega_\mathrm{s}}^{3}}} \right] 
u\\
&+1/32\,{\frac {\alpha\,\gamma\,{v}^{5}}{{\omega_\mathrm{s}}^{2}}}+1/16\,{\frac {
\Omega\,\gamma\,{v}^{3}}{{\omega_\mathrm{s}}^{2}}}
\end{aligned}
\eeq
These second-order corrections can be further simplified to
\beq
\begin{aligned}
    \Delta u_2&=\frac{\gamma\Omega r^2}{16\omega_\mathrm{s}^2}u
    +\frac1{8\omega_\mathrm{s}}\left[\frac{11\gamma^2r^4}{32}
            -\frac{3\gamma^2r^2}{2}
            +\frac{(\Omega^2+\gamma^2\omega_\mathrm{s}^2)}{\omega_\mathrm{s}^2}\right]v\\
            &+\frac{\alpha\gamma}{32\omega_\mathrm{s}^2}r^4u+\frac{51\alpha^2}{256\omega_\mathrm{s}^3}r^4v+\frac{3\alpha\Omega}{8\omega_\mathrm{s}^3}r^2v\\
\Delta v_2&=-\frac1{8\omega_\mathrm{s}}\left[\frac{11\gamma^2r^4}{32}
                    -\frac{3\gamma^2r^2}{2}
                    +\frac{(\Omega^2+\gamma^2\omega_\mathrm{s}^2)}{\omega_\mathrm{s}^2}\right]u
            +\frac{\gamma\Omega r^2}{16\omega_\mathrm{s}^2}v\\
            &-\frac{51\alpha^2}{256\omega_\mathrm{s}^3}r^4u-\frac{3\alpha\Omega}{8\omega_\mathrm{s}^3}r^2u+\frac{\alpha\gamma}{32\omega_\mathrm{s}^2}r^4v\\
\end{aligned}
\eeq

\beq
\begin{aligned}
    \dot u &= \frac1{2
    \omega_\mathrm{s}}\left\{ \gamma \omega_\mathrm{s} \left[1- \frac{r^2}4\left(1-\frac\Omega{2\omega_\mathrm{s}^2}\right)\right]u-\left(\tilde\Omega+\frac34
    \alpha r^{2}\right) v\right\}+\delta u_2,\\
    \dot v &= \frac1{2 \omega_\mathrm{s}}\left\{\gamma\omega_\mathrm{s}\left[1-\frac{r^2}4\left(1-\frac\Omega{2\omega_\mathrm{s}^2}\right)\right]v
    +\left(\tilde\Omega + \frac34\alpha r^2\right)u\right\}
    +\delta v_2
    ,
\end{aligned}
\label{avg2VDPolDuff}
\eeq
where
\[
\begin{aligned}
    \delta u_2 &=\frac{\alpha\gamma}{32\omega_\mathrm{s}^2}r^4u+\frac{51\alpha^2}{256\omega_\mathrm{s}^3}r^4v+\frac{3\alpha\Omega}{8\omega_\mathrm{s}^3}r^2v\\
    \delta v_2 &=-\frac{51\alpha^2}{256\omega_\mathrm{s}^3}r^4u-\frac{3\alpha\Omega}{8\omega_\mathrm{s}^3}r^2u+\frac{\alpha\gamma}{32\omega_\mathrm{s}^2}r^4v
\end{aligned}
\]
Incorporating these corrections explicitly in the second-order
averaged system of equations, we find
\beq
\begin{aligned}
    \dot u &= \frac1{2
    \omega_\mathrm{s}}\left\{ \gamma \omega_\mathrm{s} \left[1- \frac{r^2}4\left(1-\frac\Omega{2\omega_\mathrm{s}^2}\right)+\frac{\alpha r^4}{16\omega_\mathrm{s}^2}\right]u-\left(\tilde\Omega+\frac34
    \tilde\alpha r^{2}-\frac{51\alpha^2r^4}{128\omega_\mathrm{s}^2}\right) v\right\},\\
    \dot v &= \frac1{2 \omega_\mathrm{s}}\left\{\left(\tilde\Omega + \frac34\tilde\alpha r^2-\frac{51\alpha^2r^4}{128\omega_\mathrm{s}^2}\right)u+\gamma\omega_\mathrm{s}\left[1-\frac{r^2}4\left(1-\frac\Omega{2\omega_\mathrm{s}^2}\right)+\frac{\alpha r^4}{16\omega_\mathrm{s}^2}\right]v
    \right\}
    ,
\end{aligned}
\label{avg2VDPolDuff2}
\eeq
where $\tilde\alpha=\alpha(1-\Omega/\omega_\mathrm{s}^2)$.
From the characteristic equation, we can find the amplitude and
the frequency of the self-sustained oscillations from these equations
\beq
\begin{aligned}
    1- \frac{\bar r^2}4\left(1-\frac\Omega{2\omega_\mathrm{s}^2}\right)+\frac{\alpha \bar r^4}{16\omega_\mathrm{s}^2} &= 0\\
    \tilde\Omega+\frac34
    \tilde\alpha \bar r^{2}-\frac{51\alpha^2\bar r^4}{128\omega_\mathrm{s}^2} &=0
\end{aligned}
\eeq
One can then solve for $\omega_\mathrm{s}$ and then find $\bar r$.
\section{Averaging to second order: Forced van der Pol-Duffing oscillator}

\beq
\begin{aligned}
    \dot u &= \frac1{2
    \omega_\mathrm{d}}\left[ \gamma \omega_\mathrm{d} \left(1- \frac{r^2}4\right)u-\left(\Omega+\frac34\alpha r^{2}\right) v\right]+\Delta u_2,\\
    \dot v &= \frac1{2 \omega_\mathrm{d}}\left[\gamma\omega_\mathrm{d}\left(1-\frac{r^2}4\right)v
    +\left(\Omega + \frac34\alpha r^2\right)u-F_\mathrm{d}\right]+\Delta v_2,
\end{aligned}
\label{avg2Eqs2}
\eeq
\beq
\begin{aligned}
    \Delta u_2&={\frac {\alpha\gamma\,{u}^{5}}{32\omega_\mathrm{d}^2}}
    +\frac{\left(11\gamma^2\omega_\mathrm{d}^2+51\alpha^2\right) vu^4}{256\omega_\mathrm{d}^3}
+\gamma\left(\alpha v^2+\Omega
 \right)\frac{u^3}{16\omega_\mathrm{d}^2}\\
 &-\left[\left(-22\gamma^2\omega_\mathrm{d}^2-102\alpha^2 \right)v^3
+\left( -96\,\alpha\,\Omega+48\,{\gamma}^{2}\omega_\mathrm{d}^2 \right) v
+8F_d\gamma\omega_\mathrm{d} \right]\frac{u^2}{256\omega_\mathrm{d}^3}\\
&+
 \left( \frac {\alpha\,\gamma\,v^4}{2}+
\gamma\,\Omega\,v^2
-\frac{3F_d}{\omega_\mathrm{d}}\alpha v \right)\frac{u}{16\omega_\mathrm{d}^2}\\
&+\frac{\left(11\gamma^2\omega_\mathrm{d}^2+51\,{\alpha}^{2} \right) {v}^{5}}{256\omega_\mathrm{d}^3}
-\frac{3\left(-2\alpha\Omega+\gamma^2\omega_\mathrm{d}^2\right) {v}^{3}}{16\omega_\mathrm{d}^3}\\
&+\frac{5F_d\gamma v^2}{32\omega_\mathrm{d}^2}
+\frac {\left( \Omega^2+\gamma^2\omega_\mathrm{d}^2\right)v}{8\omega_\mathrm{d}^3}
-\frac{F_\mathrm{d}\gamma}{8\omega_\mathrm{d}^2},\\
\Delta v_2 &=-\,{\frac { \left( 51\,{\alpha}^{2}+11\,{\gamma}^{2}{
\omega_\mathrm{d}}^{2} \right) {u}^{5}}{{256\omega_\mathrm{d}}^{3}}}+\frac {\alpha\gamma v
u^4}{32\omega_\mathrm{d}^2}\\
&-\left[\left( 102\alpha^2+22\gamma^2\omega_\mathrm{d}^2\right)v^2
+96\alpha\Omega-48\gamma^2\omega_\mathrm{d}^2\right]\frac{u^3}{256\omega_\mathrm{d}^3}\\
& + \left( \,{
\frac {\alpha\,\gamma\,{v}^{3}}{{\omega_\mathrm{d}}^{2}}}+\,{\frac {\Omega\,
\gamma\,v}{{\omega_\mathrm{d}}^{2}}}+\frac {9F_d
\alpha}{2\omega_\mathrm{d}^3} \right) \frac{u^2}{16}\\
 -&\left[\left(51\alpha^2+11\gamma^2
\omega_\mathrm{d}^2 \right) v^4+
 48\left(2\alpha\Omega-\gamma^2\omega_\mathrm{d}^2 \right) v^2
+48F_d\gamma\omega_\mathrm{d} v
+32(\gamma^2\omega_\mathrm{d}^2+\Omega^2) \right] 
\frac{u}{256\omega_\mathrm{d}^3}\\
&+\frac{\alpha\gamma v^5}{32\omega_\mathrm{d}^2}+\frac{
\Omega\gamma v^3}{16\omega_\mathrm{d}^2}
+\frac{3F_d\alpha v^2}{32\omega_\mathrm{d}^3}+\frac{F_\mathrm{d}\Omega}{8\omega_\mathrm{d}^3}.
\end{aligned}
\eeq
We can further simplify the above equations and obtain
\beq
\begin{aligned}
    \dot u &= \frac1{2
    \omega_\mathrm{d}}\left\{ \gamma \omega_\mathrm{d} \left[1- \frac{r^2}4\left(1-\frac\Omega{2\omega_\mathrm{d}^2}\right)+\frac{\alpha r^4}{16\omega_\mathrm{d}^2}\right]u-\left(\tilde\Omega+\frac34
    \tilde\alpha r^{2}-\frac{51\alpha^2r^4}{128\omega_\mathrm{d}^2}\right) v\right\}+\delta u_2,\\
    \dot v &= \frac1{2 \omega_\mathrm{d}}\left\{\left(\tilde\Omega + \frac34\tilde\alpha r^2-\frac{51\alpha^2r^4}{128\omega_\mathrm{d}^2}\right)u+\gamma\omega_\mathrm{d}\left[1-\frac{r^2}4\left(1-\frac\Omega{2\omega_\mathrm{d}^2}\right)+\frac{\alpha r^4}{16\omega_\mathrm{d}^2}\right]v-F_\mathrm{d}\right\}+\delta v_2
    ,
\end{aligned}
\label{eq:avg2forcedVDPolDuff2}
\eeq
with the shorthand notation
\beq
\begin{aligned}
\delta u_2 &=-\frac {F_\mathrm{d}}{32\omega_\mathrm{d}^3}\left[
\gamma\omega_\mathrm{d}{u}^{2}+6\alpha uv+5\gamma\omega_\mathrm{d} v^2
+4\gamma\omega_\mathrm{d}\right],\\
\delta v_2 &= 
\frac {F_\mathrm{d}}{32\omega_\mathrm{d}^3}\left[9\alpha u^2-6\gamma\omega_\mathrm{d} uv+3\alpha v^2+4\Omega \right].
\end{aligned}
\eeq


%
\end{document}